\documentclass[a4paper]{article}
\usepackage{amsfonts}
\usepackage{amssymb}
\usepackage[margin=2cm]{geometry}
\usepackage[dvips]{graphicx}

\usepackage[T1]{fontenc}
\usepackage[latin2]{inputenc}

\usepackage{amsmath}
\usepackage{bbm}

\begin{document}


\title{Remarks on CHSH inequality, Tsirelson bound , Popescu-Rohlich boxes and inner product spaces}

\author{Robert Alicki \\
  {\small
Institute of Theoretical Physics and Astrophysics, University of
Gda\'nsk,  Wita Stwosza 57, PL 80-952 Gda\'nsk, Poland
} }

\date{\today}
\maketitle

\begin{abstract}
The EPR correlation matrix admit a  Hilbert model if it can be extended to a positive full correlation matrix.
A simple proof of the corresponding Tsirelson bound  is given and the relations to PR boxes and indefinite inner product spaces are briefly discussed.

\end{abstract}

The  setting proposed by Clauser, Horn, Shimony and Holt to derive their version of Bell's inequality \cite{CHSH}
assumes the existence of four binary observables $A_1, A_2 , B_1, B_2$ with values $\pm 1$ such that any pair $(A_i, B_j)$ consists of commensurable observables. Therefore the real symmetric correlation matrix 
\begin{equation}
\mathbf{C}^{AB}_{ij} = \langle A_i B_j\rangle \ ,\ |\langle A_i B_j\rangle|\leq 1
\label{cor1}
\end{equation}
can be obtained from the experimental data over a large enough ensemble. We can think about this situation in terms
of the Heisenberg picture with a fixed initial state which determines all averages $\langle \cdot\rangle$ and observables  $A_1, A_2 , B_1, B_2$ which can occupy arbitrary "space-time" locations with the only condition - commensurability of $(A_i, B_j)$. We say that the correlation matrix is consistent with the (real) Hilbert model if there exists
a (real) hermitian positively defined $4\times 4$ \emph{full correlation matrix} $\mathbf{C}$ of the form
\begin{equation}
\mathbf{C} =\begin{pmatrix} \mathbf{C}^{AA}& \mathbf{C}^{AB}\cr
            \mathbf{C}^{AB} & \mathbf{C}^{BB}
\end{pmatrix} 
\label{fulcor}
\end{equation}
with (real) $2\times 2$ hermitian matrices
\begin{equation}
\mathbf{C}^{AA} =\begin{pmatrix} 1 & X\cr
            {\bar X} & 1
\end{pmatrix}\ ,
\mathbf{C}^{BB} =\begin{pmatrix} 1 & Y\cr
            {\bar Y} & 1
\end{pmatrix}           
\label{fulcor1}
\end{equation}
and not directly measurable (real) quantities $ X, Y$ satisfying $|X|,|Y|\leq 1 $.\\
A classical theory with observables being real functions $f_{\alpha}(x)$ on a certain "phase space" and  a  state given by a probability measure $p(x)$ provides a real Hilbert model
with the correlation functions defined by
\begin{equation}
\mathbf{C}_{\alpha\beta} = \int p(x) f_{\alpha}(x) f_{\beta}(x) dx
\label{corquant}
\end{equation}
while the quantum theory with hermitian operators $F_{\alpha}$ as observables and a state given by a Hilbert
space vector $|\psi\rangle$ (generally a purification of an arbitrary mixed state) yields
\begin{equation}
\mathbf{C}_{\alpha\beta} = \langle\psi|F_{\alpha} F_{\beta}\psi\rangle\ .
\label{corclass}
\end{equation}
Obviously, in both cases \emph{positivity of the full correlation matrix follows from  positivity of a (real) Hilbert space inner product}.\\
A useful parameter which determines the consistency of the correlation data (\ref{cor1}) with the Hilbert space model
was defined in \cite{CHSH} as
\begin{equation}
{\cal B}(\mathbf{C}^{AB}) = |\langle A_1 B_1\rangle +\langle A_1 B_2\rangle+\langle A_2 B_1\rangle -\langle A_2 B_2\rangle |
= \sqrt{2}|\mathrm{Tr}\bigl(\mathbf{C}^{AB}\mathbf{H}\bigr)|
\label{CHSH}
\end{equation}
where $\mathbf{H}$ is a real symmetric unitary  $2\times 2$ matrix ("Hadamard gate")
\begin{equation}
\mathbf{H} =\frac{1}{\sqrt{2}}\begin{pmatrix} 1& 1\cr
            1 & -1
\end{pmatrix} \ .
\label{Had}
\end{equation}
Define a pair of obviously positive $4\times 4$ matrices 
\begin{equation}
\mathbf{R}^{\pm} =\begin{pmatrix} \mathbf{1}& \pm \mathbf{H}\cr
            \pm\mathbf{H} & \mathbf{1}
\end{pmatrix} \ .
\label{Rmat}
\end{equation}
Then the positivity of the full correlation matrix (\ref{fulcor}) implies
\begin{equation}
\mathrm{Tr}\bigl(\mathbf{C}\mathbf{R}^{\pm}\bigr)\geq 0
\label{Tbound}
\end{equation}
and hence by (\ref{cor1}), (\ref{CHSH}) and (\ref{Rmat}) the \emph{Tsirelson bound} \cite{T}
\begin{equation}
{\cal B}(\mathbf{C}^{AB}) = |\langle A_1 B_1\rangle +\langle A_1 B_2\rangle+\langle A_2 B_1\rangle -\langle A_2 B_2\rangle | \leq 2\sqrt{2}
\label{Tbound1}
\end{equation}
which is a necessary condition for the existence of the Hilbert model. The bound is tight as it can be saturated by a two qubit model with a maximally entangled state.\\
{\bf Exercise:} Prove (or disprove) that the CHSH inequality ${\cal B}(\mathbf{C}^{AB})\leq 2$ is a necessary condition for the existence of a real Hilbert model.\\
According to (\ref{CHSH}) and (\ref{cor1}) the maximal value of the parameter ${\cal B}(\mathbf{C}^{AB})$ is reached for the correlation matrix of the form
\begin{equation}
\mathbf{C}^{AB} =\pm\sqrt{2} \mathbf{H}\ , {\cal B}(\pm\sqrt{2} \mathbf{H})=4
\label{cormax}
\end{equation}
which does not admit a Hilbert model.\\
This anzatz attracted a lot of attention in Quantum Information under the name of "Popescu-Rohlich box".
\par
A question arises: \emph{Can we have a reasonable theory based on spaces with indefinite inner product which can violate the Tsirelson bound?}\\

It is not such an exotic idea as a covariant formulation of quantum electrodynamics and other gauge theories exists within a framework of spaces with indefinite inner product. However, in these cases the physical states always satisfy the condition $\langle\psi|\psi\rangle >0$ and the "ghost states" with "imaginary length" appear only in the intermediate stages of the perturbation calculus. Treating "ghost states" as physical ones would lead to serious problems with probabilistic interpretation and causality.\\
It is interesting that a similar conclusion has been reached in the recent paper \cite{Z} using the language of "Alice and Bob" .

\end{document}